# The Study of Multi-Layer sTGC Test System for ATLAS Phase-I upgrade

Feng Li, Xinxin Wang, Peng Miao, Shuang Zhou, Zhilei Zhang, Tianru Geng, Shengquan Liu, and Ge Jin

*Abstract*—A completely New Small Wheel (NSW) will be constructed for ATLAS Phase-1 upgrade. Small-Strip Thin-Gap-Chamber (sTGC) will devote to the trigger function of NSW. A full-size sTGC quadruplet consists of 4 layers, and will need 4 pad Front-End-Boards and 4 strip Front-End-Boards for sTGC signals readout. The 8 boards should be readout simultaneously at a time. This paper presents the study of multi-layer sTGC test system, a FEB Driver Card (FEBDC) is designed for pFEB and sFEB boards handling. The design and test of FEBDC are described in details.

## I. Introduction

ATLAS[1,2] is one of the four experiments at Large Hadron Collider (LHC).LHC will be upgraded in the next several years aiming to new physics study. ATLAS experiment will fulfill Phase-I upgrade by 2020. The current ATLAS muon end-cap system (Small Wheel, SW) [3] will be replaced with a completely New Small Wheel (NSW). The NSW is a set of precision tracking and trigger detectors able to work at high rates with excellent real-time spatial and time resolution. The small-strip Thin Gap Chamber (sTGC) will devote to trigger function in NSW.

STGC contains pad, wire and strip readout. The pads are used through a 3-out-of-4 coincidence to identify muon tracks roughly pointing to the interaction point (IP). They are also used to define which strips need to be readout to obtain a precise measurement in the bending coordinate for the event selection. The signals from strips and pads of sTGC quadruplets will be readout by two different front-end boards (FEB) -- strip FEB (sFEB) and pad FEB (pFEB), respectively. The FEB boards are mounted on the sTGC quadruplets.

This paper presents the study of multi-layer sTGC test system, and it is named Front End Boards Driver Card (FEBDC), which has the capability of handling four sFEBs and four pFEBs simultaneously. The connection and communication between FEBDC and the eight p/sFEBs are the same as a real sTGC Readout system. So the front-end chips on FEBs can be configured by the FEBDC and the raw data for the hit events can be readout and sent back to the FEBDC. The FEBDC can readout the sTGC signals, evaluate the sTGC performance which can help optimize the production of sTGC chambers.

This work was supported by the National Natural Science Foundation of China under Grants 11461141010 and 11375179, and in part by "the Fundamental Research Funds for the Central Universities" under grant No. WK2360000005.

Feng Li, Xinxin Wang, Peng Miao, Shuang Zhou, Zhilei Zhang, Tianru Geng, Shengquan Liu, and Ge Jin are with State Key Laboratory of Particle Detection and Electronics, University of Science and Technology of China, Hefei 230026, P.R. of China. (e-mail: phonelee@ustc.edu.cn, wxx10@mail.ustc.edu.cn, mpmp@mail.ustc.edu.cn, neo@mail.ustc.edu.cn, zzlei@mail.ustc.edu.cn, gudujian@mail.ustc.edu.cn, lsqlsq@mail.ustc.edu.cn, goldjin@ustc.edu.cn ).

## II. Test System Architecture

### A. Hardware

The schematic block diagram of the FEBDC is shown in Fig. 1. The core of FEBDC is based on a Kintex-7 FPGA, which is configured by a Serial Peripheral Interface (SPI) flash. The upper eight miniSAS connectors are designed to connect and communicate with four pad FEB and four strip FEB boards each. When four pFEB and four sFEB boards are mounted on a real sTGC quadruplet, the miniSAS connector is used to receive the configuration data from the upper driver card and send out the raw data of front-end circuits. So the FEBDC will act as the upper driver card. It will configure and collect the signal of a full scale 4-layer sTGC chamber.

The Ethernet PHY is a physical layer device for 1000BASE-T, 100BASE-TX, and 10BASE-T application. The Graphical User Interface running on a PC will download the command and configuration data to the FPGA through the Ethernet chip. FPGA will decode the command, then distributes the configuration data the corresponding FEBs through miniSAS connectors. And also the hit event raw data was readout through the miniSAS connectors. The photo of FEBDC is shown in Fig. 2

### B. Initialization and Data Readout of FEBDC

The VMM3[4,5] chips on the p/sFEB boards need to be configured, the configuration stream is 1728-bit length in one VMM3. On sFEB, all the 8 VMM3s are inter-connected in a daisy-chain, the total configuration stream will be 1728*8bits length. And on pFEB, all the three VMMs will need 1728*3bits configuration data. The configuration bits are set in a graphical user interface (GUI). The GUI is designed based on Qt platform. The GUI running on a PC will download the command and configuration data to the FPGA through the Ethernet chip. FPGA will decode the command, then works in corresponding modes. All the data from p/sFEB boards will be packaged and transmitted to PC through Ethernet also.

The GUI will also decode the event raw data and give out the hit channel number, the BCID, the amplitude and the hit time information

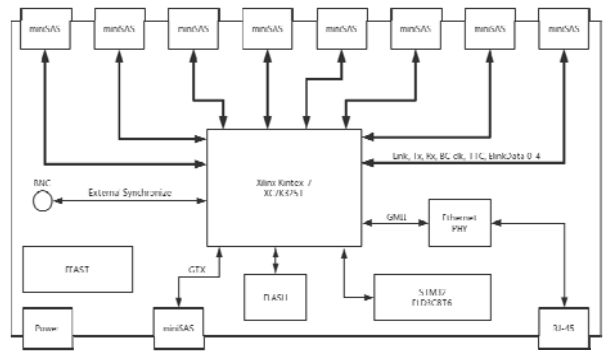

Fig 1. Block diagram of the FEBDC

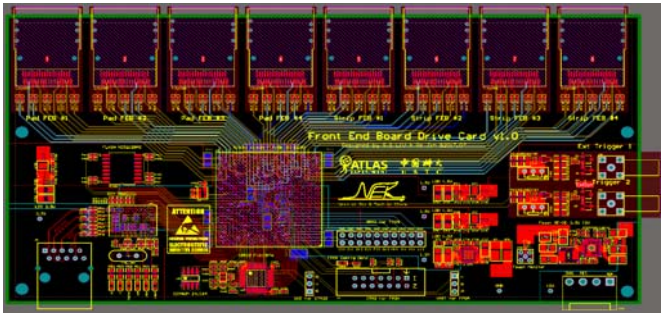

Fig 2. The Photo of FEBDC

of the event, which will help to evaluate the performance of sTGC quadruplets.

### III. Test Results

The FEBDC and p/sFEB v2.1 boards are used at Shandong University and Weizmann Institute of Israel, which are both the sTGC production sites, to study and monitor the sTGC quadruplet performance, which are shown in Fig 3 and Fig 4. They are also used for the sTGC beam test at CERN, which is shown in Fig 5. The typical spatial resolution of sTGC strips at cosmic ray test is 01.52mm, which is shown in Fig 6. The detection efficiency result of a single layer sTGC is shown in Fig 7, which means the FEBDC integration with p/sFEB boards have the ability to monitor the sTGC performance.

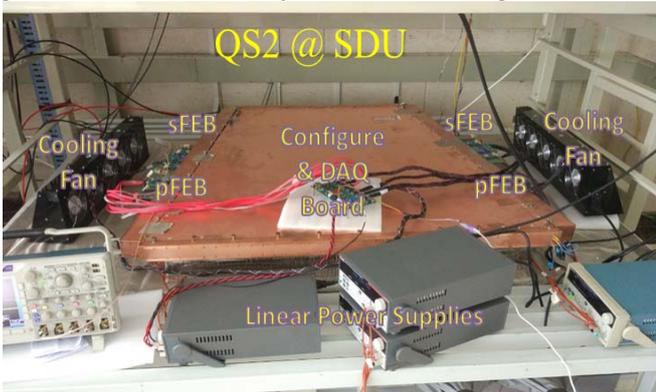

Fig 3. sTGC test system at ShanDong University, China

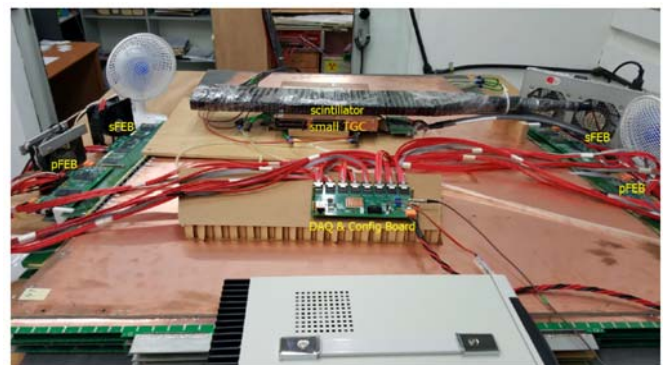

Fig 4. sTGC test system at Weizmann Institute of Israel

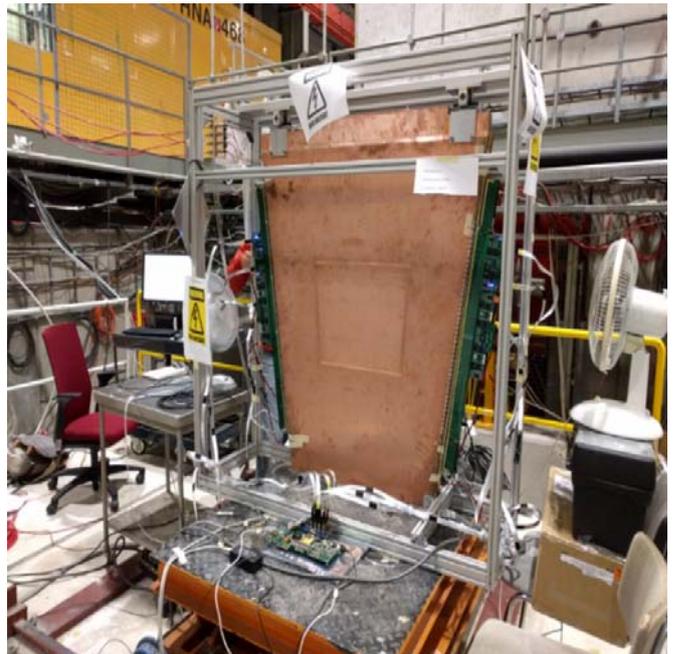

Fig 5. sTGC Beam test at CERN

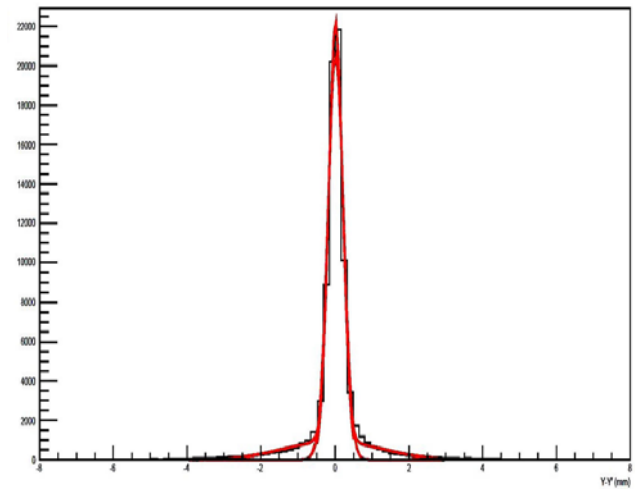

Fig 6. Typical spatial resolution result (0.152mm) for single layer sTGC at cosmic ray test

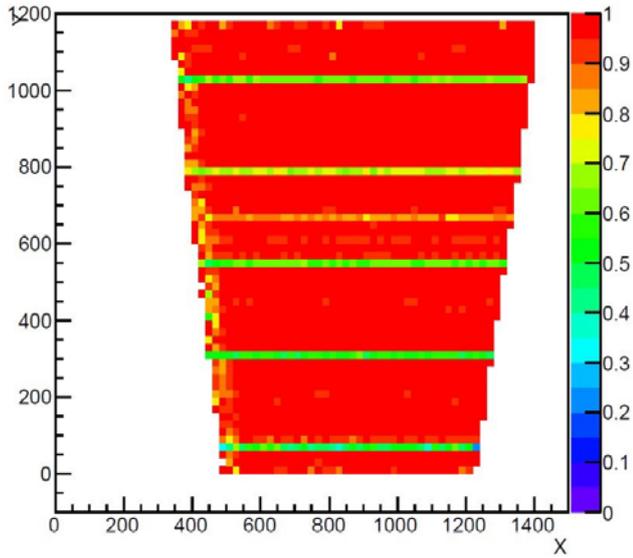

Fig 7. detection efficiency result of single layer sTGC

## IV. Conclusions

In this paper, the Front End Board Driver Card based on high performance FPGA and Gigabit Ethernet interface is described. The hardware design and data readout are discussed in details. The FEBDC has be used for sTGC chamber performance site test. And it will devote to the sTGC performance monitoring during the chamber production and evaluation.

Acknowledgment

We want to thank Brookhaven National Laboratory for the VMM3 configuration support.